\documentclass[12pt,usegraphicx,notitlepage,natbib]{article}
\pdfoutput=1
\usepackage{graphicx}
\usepackage{placeins}
\usepackage[usenames]{color}
\definecolor{dblue}{rgb}{0.05,0.05,0.35}
\definecolor{grey}{rgb}{0.35,0.35,0.35}
\usepackage[colorlinks=true,breaklinks=true,bookmarks=true,hypertexnames=false,bookmarksnumbered=true,backref=page,pagebackref=true,citecolor= dblue,linkcolor=grey,bookmarksopen=false,bookmarks=true,bookmarksnumbered=true]{hyperref} 
\usepackage[paper=a4paper,margin=0.5in]{geometry}% http://ctan.org/pkg/geometry
\usepackage{titlepic}
\usepackage[square,sort,comma,numbers,authoryear]{natbib}
\usepackage{subfigure}

\begin{document}
%\begin{titlepage}
%\thispagestyle{empty}
\title{DisPerSE: robust structure identification in 2D and 3D}
%\author[]{\ni Thierry Sousbie\thanks{{tsousbie@gmail.com}}}
\author{Thierry Sousbie\thanks{contact: {sousbie@iap.fr}, {tsousbie@gmail.com}}}
\date{}
%\titlepic{\includegraphics[width=\textwidth]{logo_disperse_small.jpg}}
\maketitle
\centering\includegraphics[width=\textwidth]{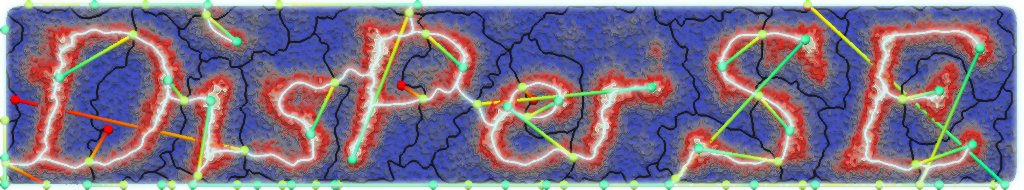}

%\vspace*{\fill}
\abstract{
We present the DIScrete PERsistent Structures Extractor (DisPerSE), an open source software for the automatic and robust identification of structures in 2D and 3D noisy data sets. The software is designed to identify all sorts of topological structures, such as voids, peaks, sources, walls and filaments through segmentation, with a special emphasis put on the later ones. Based on discrete Morse theory, DisPerSE is able to deal directly with noisy datasets using the concept of persistence (a measure of the robustness of topological features) and can be applied indifferently to various sorts of data-sets defined over a possibly bounded manifold : 2D and 3D images, structured and unstructured grids, discrete point samples via the delaunay tesselation, Healpix tesselations of the sphere, \ldots\newline

Although it was initially developed with cosmology in mind, various I/O formats have been implemented and the current version is quite versatile. It should therefore be useful for any application where a robust structure identification is required as well as for studying the topology of sampled functions (e.g. computing persistent Betti numbers).\newline

DisPerSE can be downloaded directly from the website \url{http://www2.iap.fr/users/sousbie/} and a thorough online documentation is also available at the same address.
}
%\vspace*{\fill}

%\end{abstract}

%\end{titlepage}
\newpage
\section{Basic principles}
In this section, we briefly present an overview of the mathematical concepts used in DisPerSE. For a much more detailed description of the actual implementation, see \citet{paper1} and \citet{paper2}.
\subsection{Morse theory}
\begin{figure*}
\begin{minipage}[c]{\linewidth}
\includegraphics[width=\linewidth]{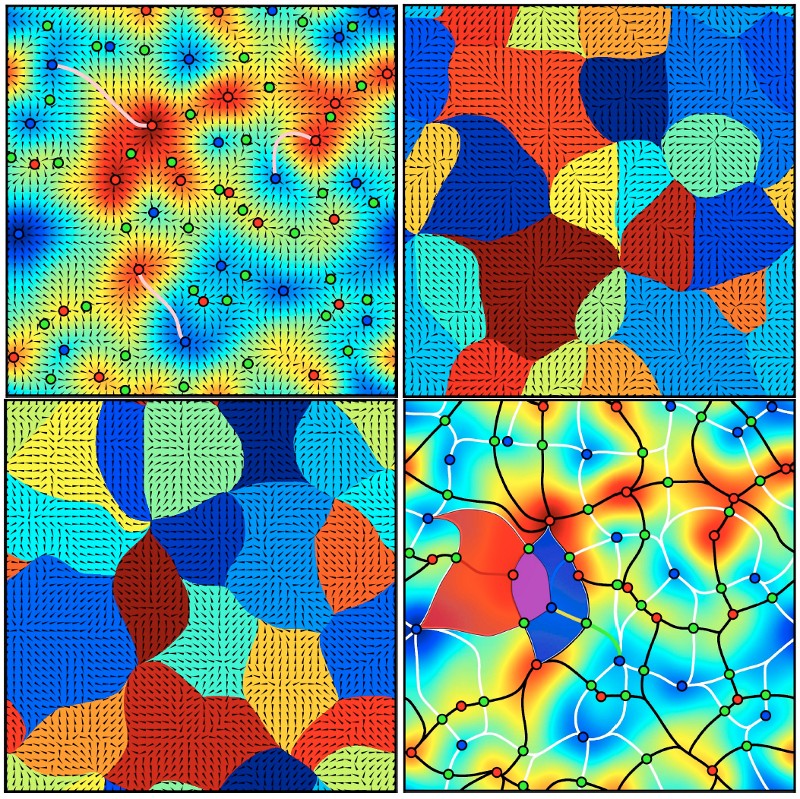}
\caption{{\em Upper-left:} critical points (minima, saddle points and maxima pictured as blue, green and red disks), and three integral lines (pink curves) of a Morse function. Black arrows show the gradient of that function. {\em Upper right:} ascending 2-manifolds : the set of points belonging to integral lines whose destination is the same minimum (critical point of index 0). {\em Lower left:} descending 2-manifolds : the set of points belonging to integral lines whose origin is the same maximum (critical point of index 2). {\em Lower right:} the Morse-Smale complex : a natural tesselation of space into cells induced by the gradient fo the function. Each cell is the set of points belonging to integral lines whose origin and destination are identical (i.e. each cell is the intersection of an ascending and a descending manifold). The purple region is a 2-cell: intersection of an ascending and a descending 2-manifold (red and blue regions) where all field lines have the same orgin and destination (a minimum and a maxium). The yellow curve is a 1-cell (also called an arc): the intersection of and ascending 2-manifold (blue region) and a descending 1-manifolds (green+yellow curves, originating from the same saddle point).\label{morse_ill}}
\end{minipage}
\end{figure*}

In DisPerSE, structures are identified as components of the Morse-Smale complex of an input function defined over a - possibly bounded - manifold. The Morse-Smale complex of a real valued so-called Morse function is a construction of Morse theory which captures the relationship between the gradient of the function, its topology, and the topology of the manifold it is defined over. Two central notions in Morse theory are that of {\em critical point} and {\em integral line} (also called {\em field line}). They are illustrated on figure \ref{morse_ill} and can be roughly described as follows:
\begin{description}
\item[Critical points] are the discrete set of points where the gradient of the function is null. For a function defined over a 2D space, there are three types of critical points (4 in 3D, ...), classified by their critical index. In 2D, minima have a critical index of 0, saddle points have a critical index of 1 and maxima have a critical index of 2. In 3D and more, different types of saddle points exist, one for each non extremal critical index. 
\item[Integral lines] are curves tangent to the gradient field in every point. There exist exactly one integral line going through every non critical point of the domain of definition, and gradient lines must start and end at critical points (i.e. where the gradient is null).\newline
\end{description}

Because integral lines cover all space (there is exactly one critical line going through every point of space) and their extremities are critical points, they induce a tessellation of space into regions called ascending (resp. descending) $k$-manifolds where all the field lines originate (respectively lead) from the same critical point (see ascending and descending $2$-manifolds on figure \ref{morse_ill}, upper right and lower left panels). The number of dimensions $k$ of the regions spanned by a k-manifold depend directly on the critical index of the corresponding critical point: descending $k$-manifolds originate from critical points of critical index $k$ while the critical index is $N-k$ for ascending manifolds, with $N$ the dimension of space.\newline

The set of all ascending (or descending) manifolds is called the {\em Morse complex} of the function. The {\em Morse-Smale complex} is an extension of this concept: the tessellation of space into regions called $p$-cells where all the integral lines have the same origin and destination (see figure 1, lower right frame). Each $p$-cell of the Morse-Smale complex is the intersection of an ascending and a descending manifold and the Morse-Smale complex itself is a natural tessellation of space induced by the gradient of the function. Figure \ref{morse_str} below illustrates how components of the Morse-Smale complex can be used to identify structures in a 3D distribution.

\begin{figure*}
\begin{minipage}[c]{\linewidth}
\includegraphics[width=\linewidth]{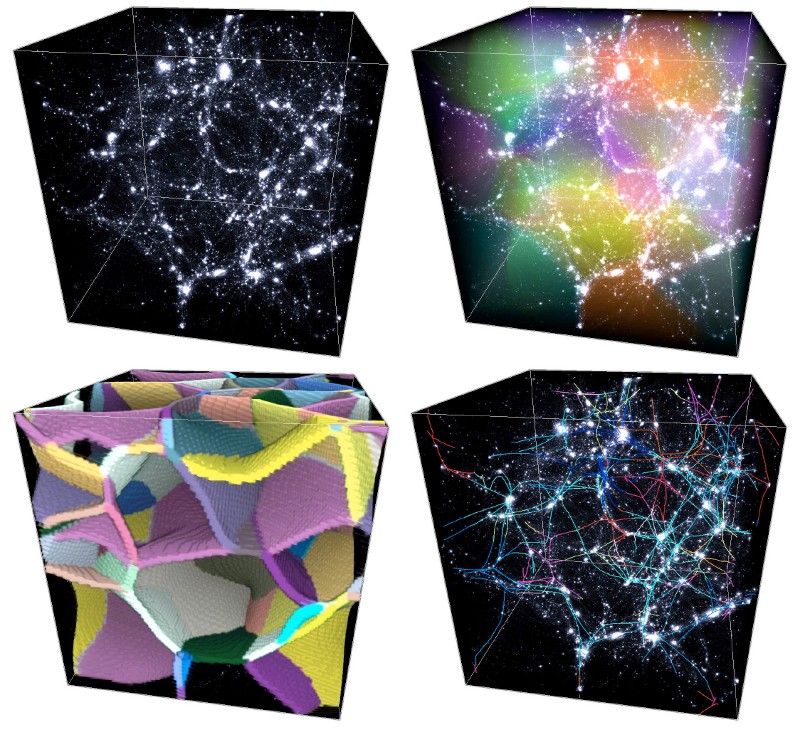}
\caption{3D structures identified as component of the Morse-Smale complex.{\em Upper-left:} density distribution of dark matter in a chunk of the Universe represented by tracer particles from a N-Body cosmological simulation.{\em Upper-right:} ascending 3-manifolds tracing the voids{\em Lower-left:} ascending 2-manifolds tracing the walls. {\em Lower-right:} the set of arcs with one maximum at their extremity, also called upper skeleton, tracing the filamentary structures. The maxima, not represented here, identify dark matter halos onto which filaments plug.\label{morse_str}}
\end{minipage}
\end{figure*}
\FloatBarrier

\subsection{Persistence and simplification}
\begin{figure*}
\begin{minipage}[c]{\linewidth}
\centering \includegraphics[width=0.8\linewidth]{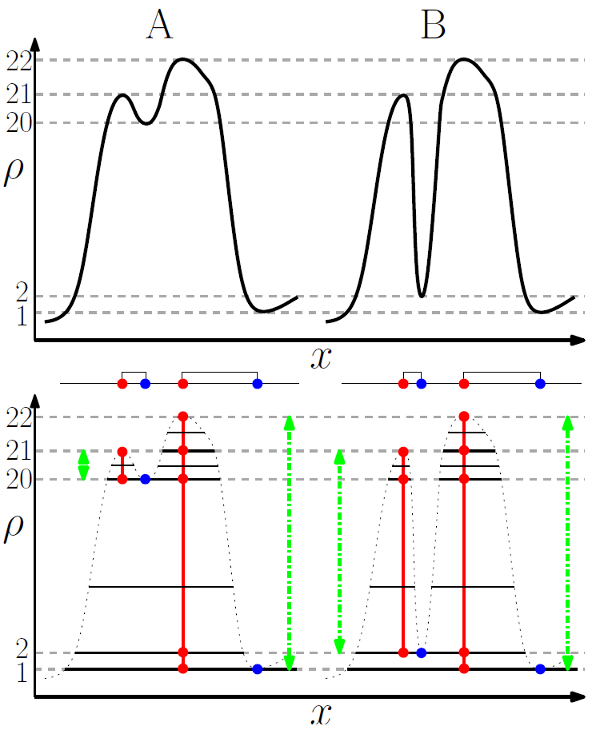}
\caption{Persistence pairs identification. {\em Top:} Two functions with similar topology (i.e. two peaks). {\em Bottom:} Changes in topology of excursion sets (sets of point with value lower than a decreasing threshold, see main text above). In 1D, new components are created at maxima and destroyed at minima. Two persistence pairs represented by green lines account for the two peaks in each function. The length of the green lines corresponds to the persistence of the pairs, which correctly accounts for the the fact that A has a small bump on the top of a peak (one high and one low persistence pair), while B has two peaks (two high persistence pairs).\label{persistence}}
\end{minipage}
\end{figure*}

Persistence itself is a relatively simple but powerful concept. To study the topology of a function, one can measure how the topology of its excursion sets (i.e. the set of points with value higher than a given threshold) evolves when the threshold is continuously and monotonically changing. Whenever the threshold crosses the value of a critical point, the topology of the excursion change. Supposing that the threshold is sweeping the values of a 1D function from high to low, whenever it crosses the value of a maximum, a new component appears in the excursion, while two components merge (i.e. one is destroyed) whenever the threshold crosses the value of a minimum. This concept can be extended to higher dimensions (i.e. creation/destruction of hole, spherical shells, ....) and in general, whenever a topological component is created at a critical point, the critical point is labeled positive, while it is labeled negative if it destroys a topological component. Using this definition, topological components of a function can be represented by pairs of positive and negative critical points called persistence pairs. The absolute difference of the value of the critical points in a pair is called its persistence : it represents the lifetime of the corresponding topological component within the excursion set (see figure \ref{persistence}).

The concept of persistence is powerful because it yields a simple way to measure how robust topological components are to local modifications of a function values. Indeed, noise can only affect a function's topology by creating or destroying topological components of persistence lower that its local amplitude. Therefore, it suffice to know the amplitude of noise to decide which components certainly belong to an underlying function and which may have been affected (i.e. created or destroyed) by noise. In DisPerSE, a persistence threshold can be specified (see options ''-nsig'' and ''-cut'' of the \href{http://www2.iap.fr/users/sousbie/web/html/index4656.html?post/mse}{mse program}) to remove topological components with persistence lower than the threshold and therefore filter noise from the Morse-Smale complex (see figure \ref{simplification} below).
\begin{figure*}
\begin{minipage}[c]{\linewidth}
\centering\includegraphics[width=0.8\linewidth]{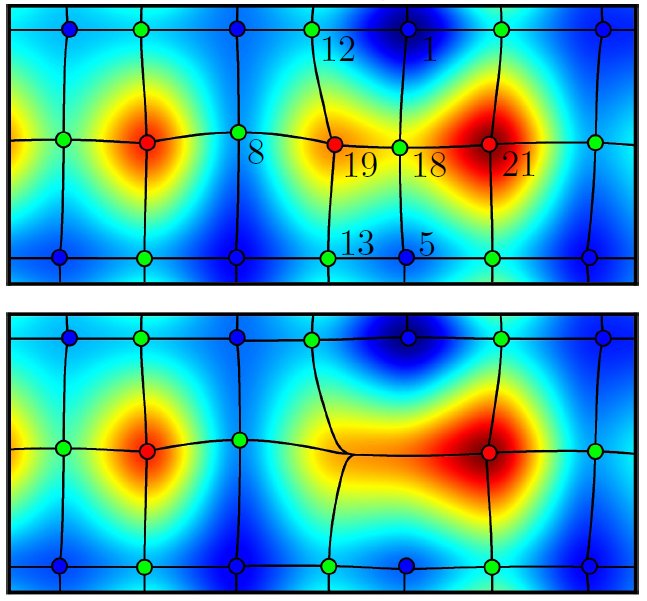}
\caption{Simplification of a low persistence pair (formed by maximum 19 and saddle point 18 on the top frame). Disks represent critical points (minima, saddle-points and maxima in blue, green and red respectively) and black lines show the arcs of the Morse-Smale complex. When the peak 19 identified by the pair is removed (lower frame), the Morse-Smale reconnects in a predetermined way that depends only on its structure.\label{simplification}}
\end{minipage}
\end{figure*}

A very useful way to set the persistence threshold is to plot a persistence diagram, in which all the persistence pairs are represented by points with coordinates the value at the critical points in the pair (see the {\em tutorial} section of the documentation to learn how to compute, plot and use persistence diagrams).
\FloatBarrier
\section{Examples of applications}
We present here a few applications to astrophysical problems, for discretely sampled 2D and 3D fields, 2D images and a function defined over a sphere. Note that DisPerSE can also be directly applied to 3D images and functions defined over arbitrary unstructured networks. A \href{http://www2.iap.fr/users/sousbie/web/html/index55a0.html?category/Quick-start}{tutorial} explaining how the structures in these exemples have been identified is available on \href{http://www2.iap.fr/users/sousbie/web/html/indexd41d.html}{DisPerSE website}.
\begin{figure*}
\begin{minipage}[c][\textheight]{\linewidth}
\begin{centering}
\subfigure[A 2D slice of a discretely sampled 3D dark matter distribution in a numerical simulation of a 50Mpc chunk of the Universe. The filamentary structure of the cosmic web is clearly visible.]{\centering \includegraphics[width=0.49\linewidth]{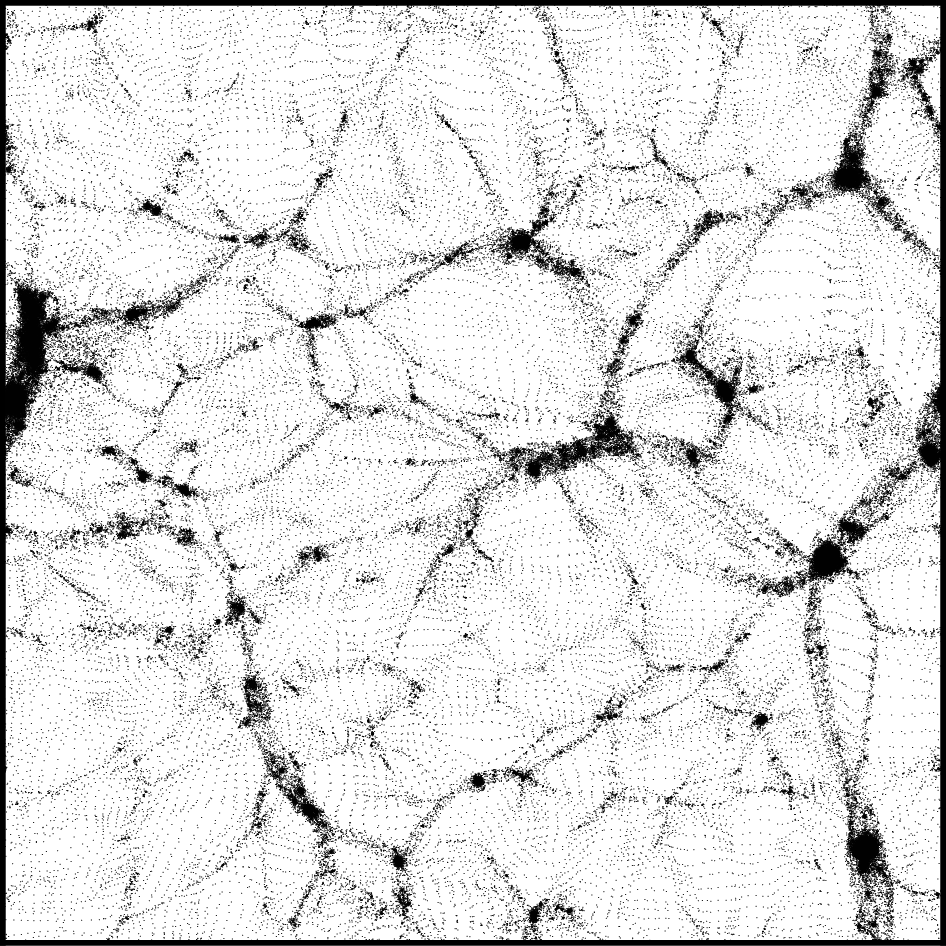}}
\subfigure[Identified filamentary structure and corresponding critical points (maxima in red, minima in blue and saddle points in yellow) ]{\centering \includegraphics[width=0.49\linewidth]{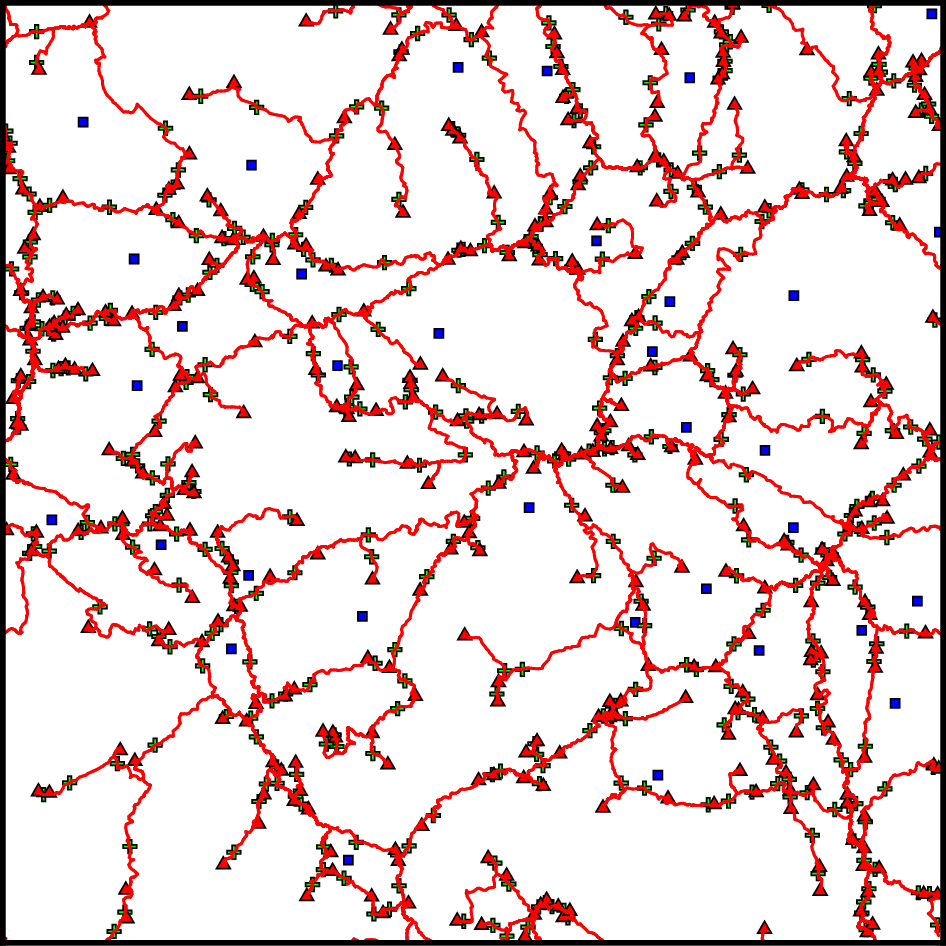}}\newline
\subfigure[Zoom on a clump, the color corresponds to the density computed by DTFE]{\centering \includegraphics[width=0.49\linewidth]{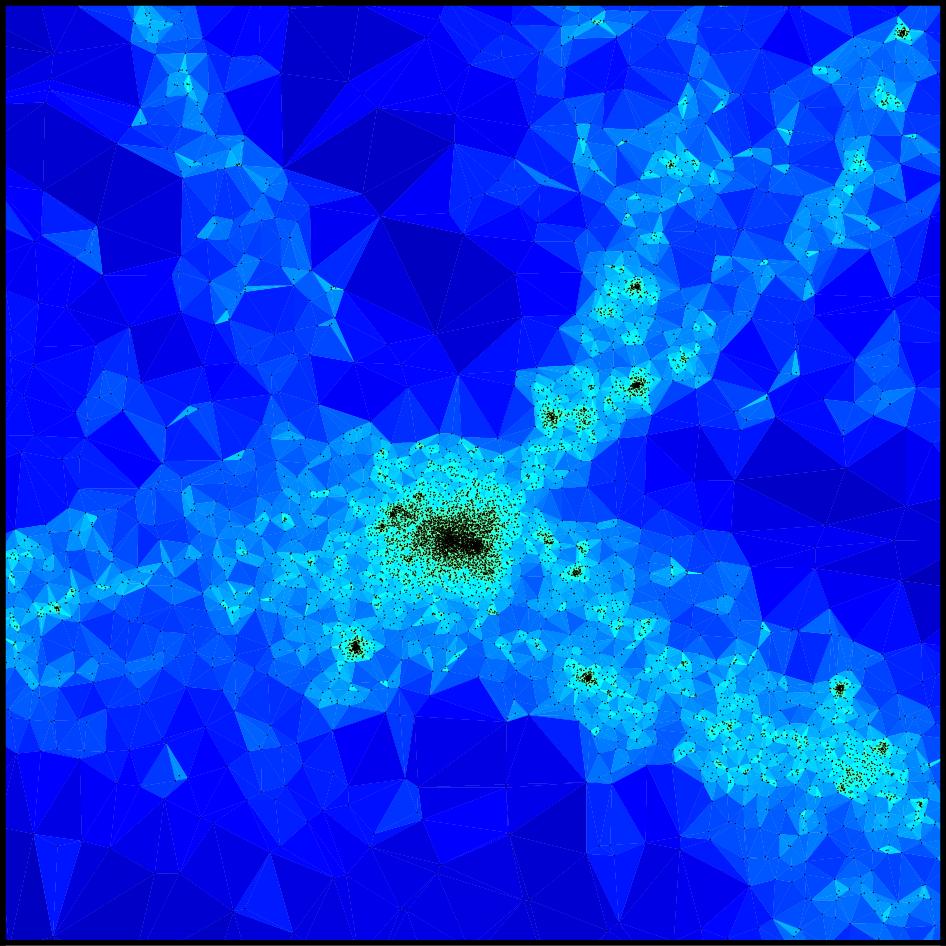}}
\subfigure[Persistent filaments connecting to a clump, note how maxima (red triangles) correctly identify bound structures]{\centering \includegraphics[width=0.49\linewidth]{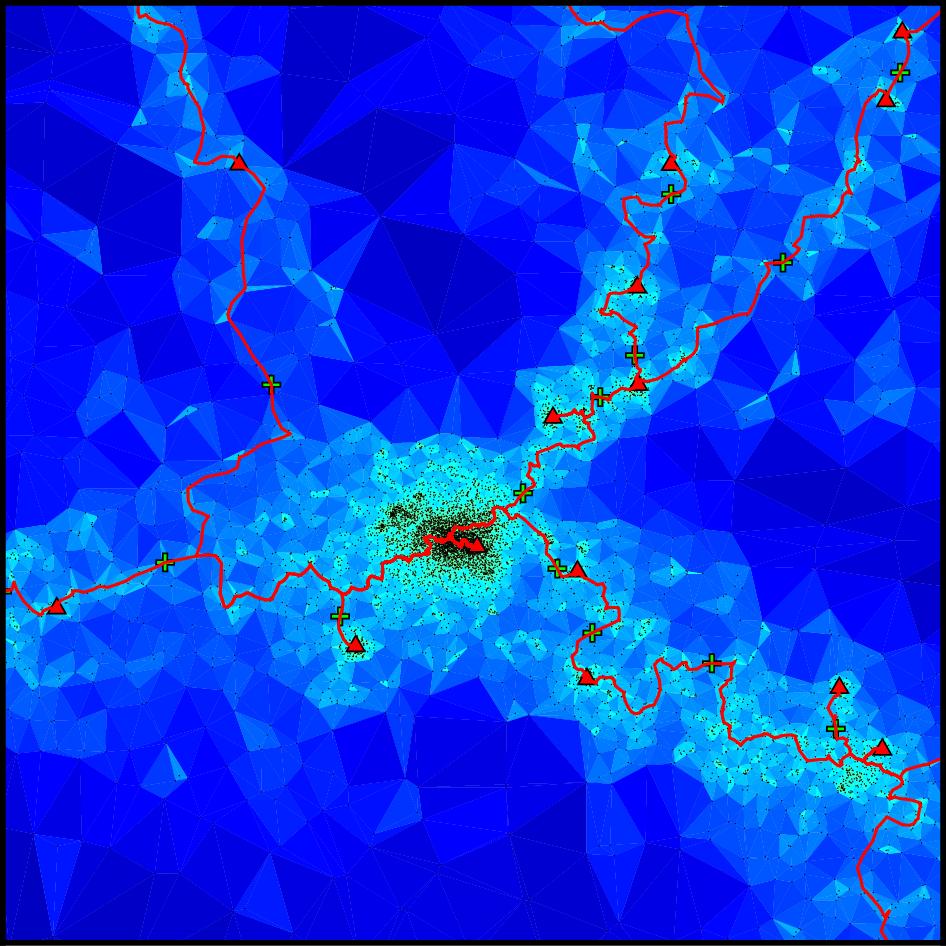}}
\end{centering}
\caption{Identification of filamentary structures and critical points from on a discretely sampled distribution using DisPerSE.}
\end{minipage}
\end{figure*}

\begin{figure*}
\begin{minipage}[c][0.49\textheight]{\linewidth}
\begin{centering}
\subfigure[Filamentary structures and a void (lower right) identified in the simulated distribution of dark matter. Structures where identified over the full 3D distribution, but only a slice is shown here.]{\centering \includegraphics[width=0.49\linewidth]{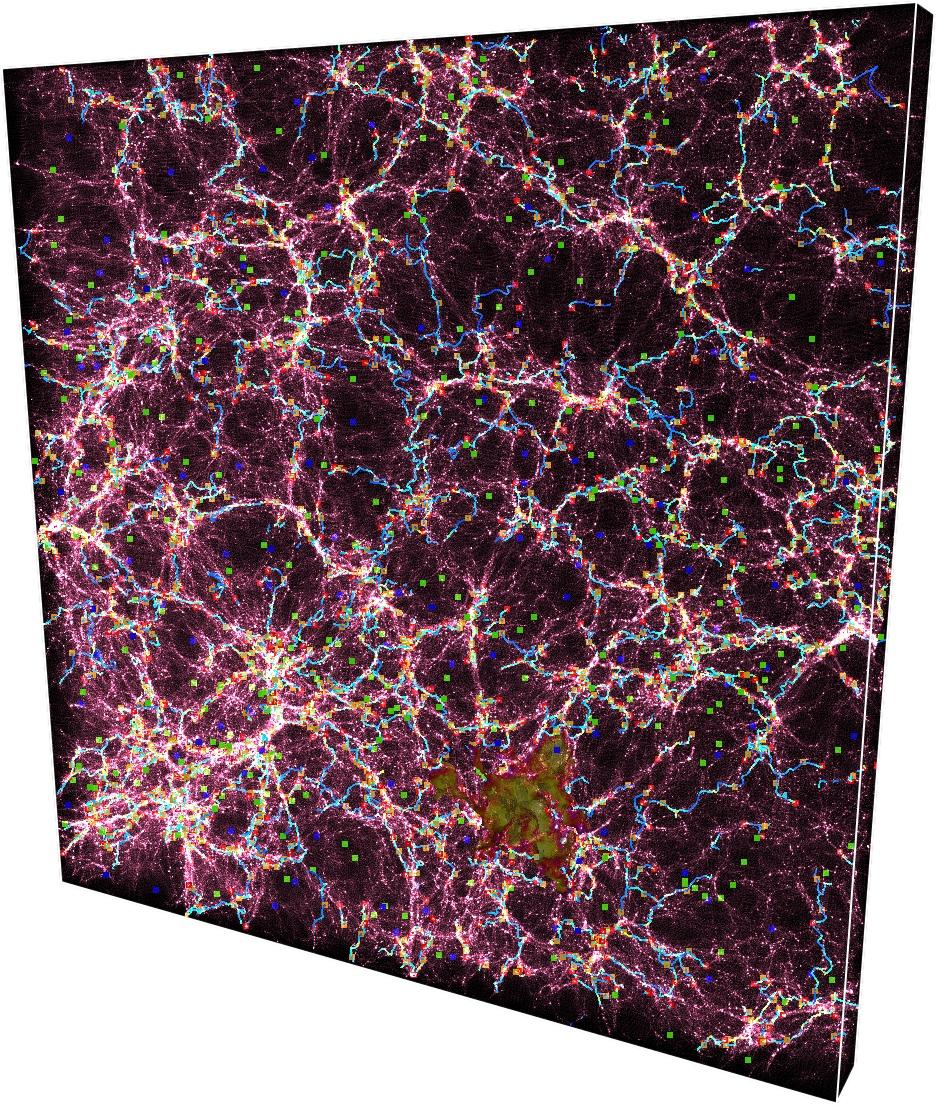}}
\subfigure[Zoom on one of the void (i.e. low density) regions, identified as a persistent ascending 3-manifold of the density field. Color corresponds to the density on the surface of the void. ]{\centering \includegraphics[width=0.49\linewidth]{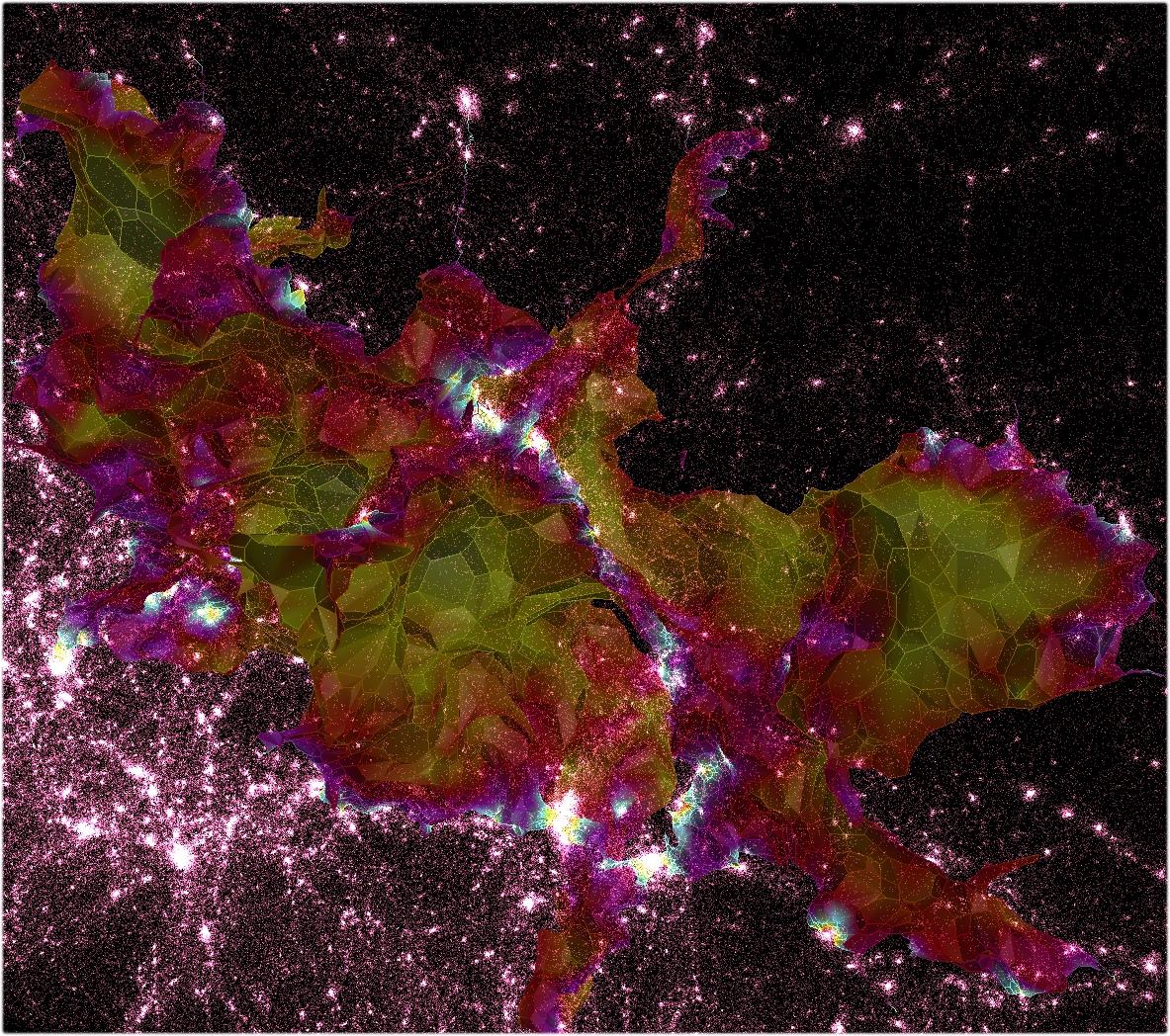}}
\caption{three dimensional filaments, voids and critical points identification in a $512^3$ dark matter particles N-body simulation}
\end{centering}
\end{minipage}
\end{figure*}

\begin{figure*}
\begin{minipage}[c][0.49\textheight]{\linewidth}
\begin{centering}
\includegraphics[width=\linewidth]{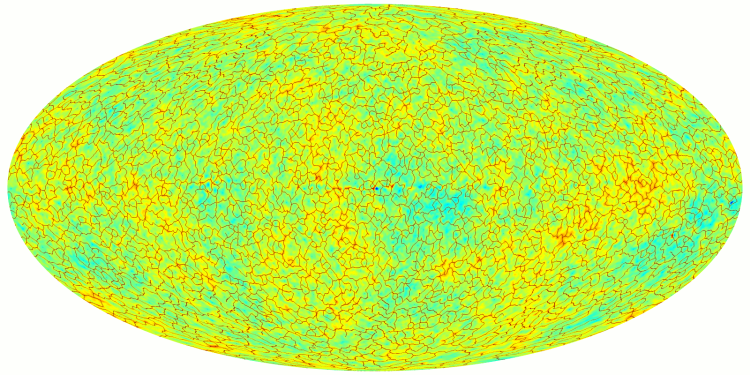}
\caption{Filamentary structures (red) in the cosmic microwave background (CMB) measured on the sky. Structures were directly identified over an Healpix type tessellation of the sphere.}
\end{centering}
\end{minipage}
\end{figure*}

\begin{figure*}
\begin{minipage}[c][0.49\textheight]{\linewidth}
\begin{centering}
\includegraphics[width=\linewidth]{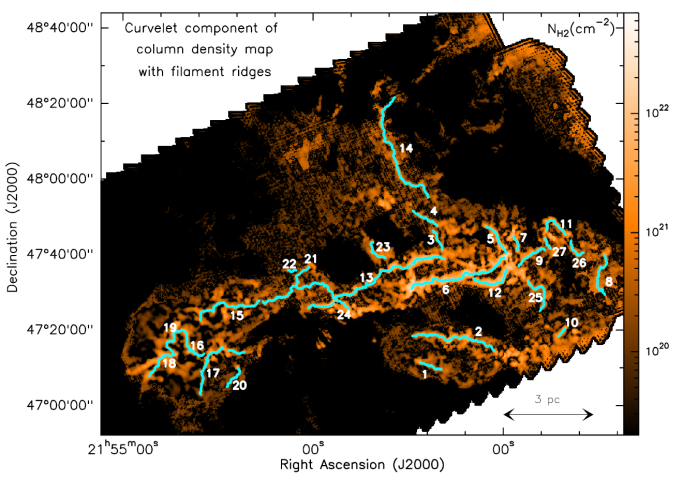}
\caption{The most prominent pieces of filaments identified with DisPerSE within a star forming region in the interstellar medium as observed by \href{http://herschel.esac.esa.int/}{HERSCHEL satellite}. The background image is a post treated version (using curvelet analysis) of the original noisy data set from which the filaments were extracted.}
\end{centering}
\end{minipage}
\end{figure*}

\bibliographystyle{natlib}

\end{document}